\documentclass[a4paper,showpacs]{revtex4}%
\usepackage{amsfonts}
\usepackage{amsmath}
\usepackage{amssymb}
\usepackage{graphicx}%
\providecommand{\U}[1]{\protect\rule{.1in}{.1in}}

\begin{document}
\title{Electric Polarization Induced by Gravity in Fat Branes}
\author{F. Dahia}
\affiliation{Dep. of Physics, Univ. Fed. da Para\'{\i}ba, Jo\~{a}o Pessoa, Para\'{\i}ba,
Brazil and Dep. of Physics, Univ. Fed. de Campina Grande, Campina Grande,
Para\'{\i}ba, Brazil.}
\author{Alex de Albuquerque Silva}
\affiliation{Dep. of Physics, Univ. Fed. da Para\'{\i}ba, Jo\~{a}o Pessoa, Para\'{\i}ba,
Brazil and Dep. of Physics, Univ. Fed. de Campina Grande, Sum\'{e},
Para\'{\i}ba, Brazil.}
\author{C. Romero}
\affiliation{Dep. of Physics, Univ. Fed. da Para\'{\i}ba, Jo\~{a}o Pessoa, Para\'{\i}ba, Brazil.}

\begin{abstract}
In the fat brane model, also known as the split fermion model, it is assumed
that leptons and baryons live in different hypersurfaces of a thick brane in
order to explain the proton stability without invoking any symmetry. It turns
out that, in the presence of a gravity source $M$, particles will see
different four-dimensional (4D) geometries and hence, from the point of view
of 4D-observers, the equivalence principle will be violated. As a consequence,
we show that a hydrogen atom in the gravitational field of $M$ will acquire a
radial electric dipole. This effect is regulated by the Hamiltonian
$H_{d}=-\mu\mathbf{A}\cdot\mathbf{\delta r}$, which is the gravitational
analog of the Stark Hamiltonian, where the electric field is replaced by the
tidal acceleration $\mathbf{A}$ due to the split of fermions in the brane and
the atomic reduced mass $\mu$ substitutes the electric charge.

\end{abstract}
\maketitle

\section{Introduction}

In the braneworld models, our four-dimensional spacetime is viewed as a
submanifold isometrically embedded in an ambient space with higher dimensions.
The basic feature of this scenario is the confinement of matter and fields in
the brane (when they have an energy lower than a certain level which is
expected to be of the order of 1TeV, at least), while gravity has access to
all dimensions \cite{rubakov,brane,rs2}. In this framework, the extra
dimensions might be much larger than Planck length. As a matter of fact, in
the RSII model it was shown that the extra dimension might even have an
infinity length without any phenomenological conflict \cite{rs2}.

A modified version of the RSII model, known as fat brane, assumes that the
brane has a thickness and that leptons and baryons live in different
hypersurfaces of the thick brane \cite{fat1}. The original motivation of this
model is to explain the stability of protons without using any symmetry. The
conservation of the baryonic number, that protects the proton from decaying,
is just a consequence of the split of fermions in the thick brane, since this
separation produces a strong suppression in couplings between quarks and
leptons. On the other hand, gauge fields have access to all the brane. Thus,
if the thickness of the brane is of TeV order then we might expect that traces
of the extra dimensions could be detected in experiments at LHC
\cite{fat1,effect}.

By virtue of the confinement, particles do not see the geometry of the whole
bulk but they feel the induced metric on the hypersurface where they live. If
there is no gravity source then all fermions see the same Minkowski spacetime.
However, under the gravitational influence of a mass $M$ in the brane, the
induced metric will be different for distinct slices. This means that leptons
and baryons will feel different geometries. From the point of view of
4D-observers (not aware of the extra dimensions), this will be seen as a
violation of the equivalence principle, since particles in the same
four-dimensional brane coordinates will feel different gravitational
accelerations. This tidal acceleration $\mathbf{A}$, due to the split of
fermions in the extra dimension, produces an internal force in a hydrogen
atom, inducing, in this way, an electric dipole in a tangential direction of
the brane. As we shall see, the Hamiltonian associated with the interaction
between the atom and the gravitational field of $M$ contains a dipole term
which has exactly the same form of the Stark Hamiltonian, $H_{d}%
=-\mu\mathbf{A}\cdot\mathbf{\delta r}$ (where $\mathbf{\delta r}$ denotes the
internal relative coordinates), in the first order of $GM$, where $G$ is
Newton's gravitational constant.

\section{Gravity in thick branes}

In the context of thick brane models, the brane is usually described as a
domain wall generated by a certain scalar field $\phi$ \cite{rubakov,thick}.
The presence of a mass $M$ (a star or a black hole) localized in the brane
will affect the domain wall solution. An exact solution for this system is not
known so far. Solutions for lower dimension (2-brane) are known \cite{emparan}
and some numeric solutions for a black hole confined to a thin 3-brane were
found recently \cite{wise}. However, in the case of a mass $M$ trapped in a
thick 3-brane, no exact solution of the Einstein equations coupled to the
scalar field equation has been found yet. However, based on the symmetry of
the problem, it is expected that a non-rotating mass should give rise to an
axisymmetric, static spacetime in five dimensions. In such spaces, as is well
known, there are coordinates in which the metric assumes the Weyl canonical
form\cite{ruth}, which can be put, by means of a convenient coordinate
transformation, in a Gaussian form adapted to the brane:%
\begin{equation}
ds^{2}=-e^{2A(r,z)}dt^{2}+e^{2B(r,z)}dr^{2}+e^{2C\left(  r,z\right)  }%
d\Omega^{2}+dz^{2},
\end{equation}
where $z=0$ gives the localization of the center of the brane. Due to
symmmetry the scalar field will depend only on the coordinates $r$ and $z$,
i.e., $\phi=\phi\left(  r,z\right)  $.

Even when $M=0$, the Einstein equations $^{\left(  5\right)  }G_{\mu\nu
}=\kappa T_{\mu\nu}^{\left(  \phi\right)  }$ coupled to the scalar field
equation $\square\phi-V^{\prime}\left(  \phi\right)  =0$ in five dimensions
are not easy to solve. Here $\kappa$ is the five-dimensional gravitational
constant, $T_{\mu\nu}^{\left(  \phi\right)  }$ is the usual energy-momentum
tensor of the scalar field subjected to the potential $V\left(  \phi\right)  $
and $c=1$. In some special situations, when the potential is conveniently
choosen, an exact solution of a self-gravitating domain wall can be obtained.
For example, taking $V\left(  \phi\right)  =\lambda/4\times\left(  \phi
^{2}-\eta^{2}\right)  ^{2}-\beta\lambda/3\eta^{2}\times\phi^{2}\left(
\phi^{2}-3\eta^{2}\right)  ^{2}$, the solution is \cite{thicksolution} :
\begin{align}
ds^{2} &  =e^{2a\left(  z\right)  }\left(  -dt^{2}+dr^{2}+r^{2}d\Omega
^{2}\right)  +dz^{2},\label{braneMinkowski}\\
2a\left(  z\right)   &  =-2\beta\ln\cosh^{2}\frac{z}{\varepsilon}-\beta
\tanh^{2}\frac{z}{\varepsilon},\\
\phi &  =\eta\tanh\frac{z}{\varepsilon},
\end{align}
where $\varepsilon^{2}=2/\lambda\eta^{2}$, $\beta=\kappa\eta^{2}/9$. This
solution can be interpreted as a regularized version of the RSII brane model.
Indeed, taking the parameter $\varepsilon$ (the thickness of the wall) equal
to zero, while keeping the condition $\varepsilon/2\beta=const.\equiv\ell$
($\ell$ defines the curvature radius of $AdS_{5}$ space) the RSII solution is
recovered\cite{thicksolution}.

Let us now consider a mass $M$ describing a body or a black hole confined in
the core of the domain wall. The presence of this gravity source certainly
will modify both the original metric and the scalar field. Considering the
amount of technical difficulty to solve this problem exactly, let us try to
employ approximation methods\cite{gio}. At large distances from $M$, where the
weak field regime is valid, the modification can be treated as a small
perturbation of the original solution. In this case, we can write%
\begin{align}
ds^{2}  &  =e^{2a}\left[  -(1+f)dt^{2}+\left(  1+m\right)  dr^{2}+r^{2}\left(
1+h\right)  d\Omega^{2}\right]  +dz^{2},\\
\phi &  =\eta\left(  \tanh\frac{z}{\varepsilon}+k\right)  ,
\end{align}
where $f,$ $m,$ $h$ and $k$ , which are functions of $r$ and $z$, give the
small corrections of the unperturbed metric and the scalar field. It may
happen that, due to the perturbation, the coordinate $z$ will not be adapted
to the level surface of the scalar field anymore. For instance, the center of
the domain wall $\left(  \phi=0\right)  $, which originally coincides with
$z=0$, is now, in the first order approximation, given by the equation
$z=-\varepsilon k\left(  r,0\right)  $. However, by a convenient coordinate
transformation, we can show that it is possible to restore a coordinate system
adapted to the center of the domain wall, i.e., coordinates in which the
center is localized at the hypersurface $z=0$. The great advantage of working
in these coordinates is the fact that \textquotedblleft initial
conditions\textquotedblright, i.e., the value that the correction functions
assume in the center of wall, can be easily established. For example, as the
center corresponds to $z=0$, then, we should have $k\left(  r,0\right)  =0$.
Another important condition can be immediately deduced based on the
expectation that the metric should be symmetric with respect to the center of
the wall. As $\partial_{z}$ is the normal vector of the hypersurface $z=0$,
then, it follows that in the center of the domain the first derivative with
respect to $z$ should be zero: $f_{z}=m_{z}=h_{z}=0.$ The remaining set of the
initial conditions, i.e, $f\left(  r,0\right)  ,m\left(  r,0\right)  ,h\left(
r,0\right)  $ and $k_{z}\left(  r,0\right)  $, can be determined by using the
thin brane solution as inspiration. In Ref. \cite{garriga}, Garriga and Tanaka
found the metric produced by a matter distribution with mass $M$ localized in
the thin brane in the first approximation order of $GM$. With the purpose to
obtain a connection with this thin brane solution, we are going to impose that
in $z=0$ our solution reproduces Garriga and Tanaka's result. This condition
allows us to determine $f\left(  r,0\right)  ,m\left(  r,0\right)  ,h\left(
r,0\right)  $. With these choices, the last condition $k_{z}\left(
r,0\right)  =0$ follows from the ($zz$)-component of the Einstein equations,
which corresponds to a constraint equation. Now by using the dynamical
components of the Einstein equations (those that involve second derivatives
with respect to $z$) to propagate the initial conditions into the bulk, we
find the metric around the center of the brane as a power series in $z$. In
the first correction order, the line element for $z<<\varepsilon$ and $r>>GM$
is given by%
\begin{align}
ds^{2}  &  =-e^{2a}\left(  1-\frac{2GM}{r}\left(  1+\frac{2\ell^{2}}{3r^{2}%
}-\frac{2\ell^{2}}{r^{4}}z^{2}\right)  \right)  dt^{2}\nonumber\\
&  +e^{2a}\left(  1+\frac{2GM}{r}\left(  1+\frac{\ell^{2}}{r^{2}}+\frac
{\ell^{2}}{r^{4}}z^{2}\right)  \right)  dr^{2}\nonumber\\
&  +e^{2a}\left(  r^{2}-\frac{3GM\ell^{2}}{r^{3}}z^{2}\right)  d\Omega
^{2}+dz^{2}, \label{metricM}%
\end{align}
and the scalar field solution is simply $\phi=\eta\tanh z/\varepsilon$ in the
first approximation order of $GM$.

\section{Electric dipole induced by gravity}

It is well known that the confinement of matter in a thick brane can be
obtained by means of a Yukawa-type interaction between the Dirac field and the
scalar field \cite{rubakov}. Under this interaction, the wave packet of a
massless Dirac field has a peak at the center of the domain wall and decay
exponentially in the extra dimension. When a non-null mass is taken into
account, the peak is shifted by a certain amount that depends on the particle
mass \cite{fat1}. Therefore, electrons and quarks will be localized in
different slices of the brane.

In this scenario, particles are in a bound state with respect to the
transversal direction, but they might be free in the parallel direction. If we
want to study the motion of a particle along the brane it is convenient to
consider a classical approach for this problem. In order to achieve this, it
is necessary first to provide a mechanism of confinement of test particles to
the brane, which may simulate classically the confinement of the matter field.
In Ref. \cite{dahia}, a particular Lagrangian based on the Yukawa interaction
was proposed to describe the particle's motion in this context. It was then
shown that the new Lagrangian has the effect of increasing the effective mass
of the particle due to the interaction with the scalar field, and this
modification is sufficient to ensure the localization of the particle. This
new Lagrangian was defined as $L^{2}=-\left(  m^{2}+h^{2}\varphi^{2}\right)
\tilde{g}_{AB}\dot{x}^{A}\dot{x}^{B}$, where $m$ is the rest mass of the
particle in a free state and $h$ is the coupling constant of the interaction.
The split of fermions in different slices can be added in our model by
redefining the Lagrangian as%
\begin{equation}
L^{2}=-\left(  m^{2}+h^{2}\left(  \varphi+\alpha m\right)  ^{2}\right)
\tilde{g}_{AB}\dot{x}^{A}\dot{x}^{B},
\end{equation}
where $\alpha$ is a new parameter related to the interaction. Calculating the
$5D$-momentum $P_{A}=\partial L/\partial\dot{x}^{A}$ of the particle, we find
$P^{A}P_{A}=-m^{2}-h^{2}\left(  \varphi+\alpha m\right)  ^{2}\equiv-m_{ef}%
^{2}$, for massive test particles $\left(  \tilde{g}_{AB}\dot{x}^{A}\dot
{x}^{B}=-1\right)  $. Then, we can verify directly that the effective mass
$m_{ef}$ is now affected by the presence of the scalar field. Of course, the
usual relation is recovered by turning off the interaction, i.e., by setting
$h=0$. It is worthy of mention that a similar kind of Lagrangian was also
employed, in a different context, to describe the interaction between test
particles and dilatonic fields \cite{kim}.

Due to the interaction with the scalar field, particles will move with a
proper acceleration $\mathcal{A}^{A}=-\Pi_{\quad}^{AC}\tilde{\nabla}_{C}%
\ln\mathcal{M}$, which is the gradient of the mass potential $\mathcal{M}%
\equiv e^{2a}m_{ef}^{2}/m^{2}$ projected by the tensor $\Pi^{AC}\equiv
\tilde{g}^{AC}+\dot{x}^{A}\dot{x}^{C}$ into the four-space orthogonal to the
particles' proper velocity $\dot{x}^{A}$.

When there is no additional gravity source ($M=0$) the metric in the bulk is
given by (\ref{braneMinkowski}) and the transversal motion decouples from the
motion in the tangential direction. In this case, the first integral of the
equation of motion in the $z$ direction can be obtained directly, also
implying that the transversal motion is bounded by the mass potential
$\mathcal{M}$ according to the equation $\mathcal{M}\dot{z}^{2}=\mathcal{E-M}%
$, where $\mathcal{E}$ is a constant related to the initial condition of the
motion. The function $\mathcal{M}$ plays the role of a confining potential and
has a stable equilibrium point when the parameters $h$ and $\alpha$ satisfy
appropriate conditions. It is important to stress here that the equilibrium
position $z_{0}$ depends on the mass of the particle. If we admit that $z_{0}$
is small in comparison with the brane thickness $\varepsilon$, then we can
show that the particle with mass $m$ is confined to a slice approximately
specified by $z_{0}=\sqrt{2}\alpha m\sqrt{\varepsilon\ell\kappa}/3.$ As we can
see, electrons are localized closer to the center than quarks. Of course we
can manipulate the Lagrangian in order to get the inverse result, namely,
quarks stuck in the center of the brane and electrons in an upper slice. For
our purpose, what is important here is that we can formulate a simple
classical model which has the essential characteristic of the fat brane model,
namely, the split of leptons and baryons in different slices of the brane. On
the other hand, in the tangent direction both fermions move freely in the same
induced four-dimensional Minkowski spacetime.

This situation changes when we consider the presence of a mass $M$ in the
brane. For the sake of simplicity, hereafter we are going to admit that quarks
are confined in the center of the brane, while electrons will be stuck in some
slice $z_{0}.$ With this choice, we can admit that almost all the mass $M$ is
localized in the center of the brane, since the baryons are stuck in that
hypersurface. It follows then that the metric will given by (\ref{metricM})
and therefore leptons and baryons will see different four-dimensional
geometries, since the induced metric depends on the value of $z.$ As a
consequence, the equivalence principle will be violated from the 4D
perspective and this fact can produce interesting phenomena in the brane as,
for example, the induction of an electric dipole in a hydrogen atom by
gravitational effects. In order to investigate this, let us consider the
motion of a particle in the spacetime with metric (\ref{metricM}). Due to the
symmetry of this spacetime, the energy $E$ and the axial angular momentum $L$
will be conserved and the particle's motion should obey the following
equations%
\begin{equation}
-m_{ef}\tilde{g}_{tt}\dot{t}=E \label{energy}%
\end{equation}%
\begin{equation}
m_{ef}\tilde{g}_{\phi\phi}\dot{\phi}=L \label{L}%
\end{equation}
\qquad\ We can also verify that $\theta=\frac{\pi}{2}$ is a solution of the
equations. On the other hand, the motion in the $z$ direction is not decoupled
from the radial motion. However, if the particle is in a circular motion or
static $\left(  r_{0}=const\right)  $ then the effect of the mass $M$ is just
to modify the equilibrium position of the particle by an amount of the order
of $GM$. Analyzing the radial motion we can see that there are stable circular
orbits for appropriate values of energy and angular momentum. Considering
this, it follows from equations (\ref{energy}) and (\ref{L}) that the angular
frequency of a particle in a circular motion of radius $r_{0}$ is given by:
\begin{equation}
\omega^{2}=\left(  \frac{\dot{\phi}}{\dot{t}}\right)  ^{2}=\frac{GM}{r_{0}%
^{3}}\left(  1+\frac{2\ell^{2}}{r_{0}^{2}}-\frac{10\ell^{2}}{r_{0}^{4}}%
z_{0}^{2}\right)  .
\end{equation}
This clearly shows that the angular frequency depends on the mass of the
particle through $z_{0}$. From the perspective of 4D observers, this mass
dependence will be seen as a violation of the equivalence principle. In fact,
in a circular orbit with the same radius $r_{0},$ a proton (stuck in the
center of the brane) will move faster than an electron.

Now if we consider a hydrogen atom orbiting the mass $M$ with a certain
angular frequency or static, then based on the previous reasoning we are led
to expect that the radius of the electron's orbit and the radius of the
proton's orbit should be different. As matter of fact, the center of the
negative charge tends to circulate in an outer orbit in comparison with the
proton's orbit, as we shall see next.

In the weak field regime, we can consider that the gravitational potential of
the mass $M$ is given by%
\begin{equation}
\varphi=-\frac{GM}{r}\left(  1+\frac{2\ell^{2}}{3r^{2}}-\frac{2\ell^{2}}%
{r^{4}}z^{2}\right)  .
\end{equation}
Here it is important to note that, based on the equation (\ref{energy}), the
warping factor can be incorporated in a redefinition of the mass of the
particle and for this reason we might define the potential $\varphi$ without
using it. Thus, the non-relativistic Hamiltonian of the hydrogen atom is%
\begin{equation}
H=\frac{P_{p}^{2}}{2m_{p}}+\frac{P_{e}^{2}}{2m_{e}}+m_{p}\varphi_{p}%
+m_{e}\varphi_{e}+U,
\end{equation}
where $U$ is the potential energy of the electric and gravitational
interaction between the proton and the electron. Introducing the coordinates
of the center of mass $\mathbf{R}$ and the relative coordinates
$\mathbf{\delta r}=\mathbf{r}_{p}-\mathbf{r}_{e}$, we can verify that $H$
contains an unusual dipole term due to the fact that electrons and protons
live in different slices of the brane. This new term $H_{d}$ has the same form
of the Stark Hamiltonian%
\begin{equation}
H_{d}=-\mu\mathbf{A}\cdot\mathbf{\delta r},
\end{equation}
where the tidal acceleration $\mathbf{A}$ between the electron and the proton
replaces the electric field. We must emphasize that $\mathbf{A}$ is the
relative gravitational acceleration between the electron and the proton when
they have the same four-dimensional brane coordinates. It is clear that the
origin of $\mathbf{A}$ is the split of fermions in the extra dimensions. It
can also be shown that
\begin{equation}
\mathbf{A}=-GM\left(  \frac{10\ell^{2}}{R^{7}}z_{0}^{2}\right)  \mathbf{R}.
\end{equation}
We should mention that, considering the procedure usually adopted to deal with
atomic systems in gravitational field \cite{parker} (which is based on the
expansion of metric around the center of the mass of the system in Fermi
coordinates), this tidal acceleration comes from the term $R_{\;\beta
\gamma\lambda;\sigma}^{\alpha}u^{\beta}u^{\gamma}s^{\lambda}s^{\sigma}$, where
$u^{\beta}$ is the proper velocity of the center of mass and $s^{\sigma
}=\left(  0,0,0,0,z_{0}\right)  $ is the separation vector (or the relative
coordinates). The semi-colon indicates the covariant derivative of the
five-dimensional Riemmann tensor $R_{\;\beta\gamma\lambda}^{\alpha}$.

The Hamiltonian $H_{d}$ will induce an electric dipole $\mathbf{p}$ in the
hydrogen atom, whose direction tends to be aligned with the tidal field. In
order to make a rough estimate of the dipole magnitude we are going to
describe the atom following a semi-classical approach. First, we admit that
the electron charge is uniformly distributed in a cloud around the proton. The
tidal acceleration inside the hydrogen atom will produce a radial separation
between the proton and the center of the negative charge, giving rise to an
electric force $\mathbf{F}$ between them. To calculate $\mathbf{F}$, we are
going to make some considerations. We begin by recalling that, according to
the fat brane model, the particle state is described by a very narrow wave
packet along the extra dimension and thus the wave function can be considered
as a delta-distribution in $z$ direction. So we can assume that the electronic
cloud is spread in a spherically symmetric 3-volume of the slice $z_{0}$. A
sphere attached to a hypersurface constitutes a kind a 3D-disk from the bulk
perspective. We are going to admit that the radius of this disk is equal to
the Bohr radius $a$. Additionally, it is reasonable to expect that the radial
separation $\delta r$ is not greater than $z_{0}$, and therefore, in our
calculations, we have to take into account that in such domain the electric
force has a 5D behavior. Considering all these assumptions, and also that
$z_{0}<<a$, we can show that $\mathbf{F}=\left(  2ke^{2}\varepsilon/\pi
^{2}a^{4}\right)  \mathbf{\delta r}$, where $k$ is the known electrostatic
constant in 4D spacetime.

Since at the equilibrium state the internal electric force and the tidal force
are balanced, it follows that the electric dipole induced by gravity is
proportional to the tidal acceleration $\mathbf{p}=\left(  \pi^{2}a^{4}%
\mu/2ke\varepsilon\right)  \mathbf{A}.$ The proportionality factor defines the
electric polarizability of the hydrogen atom induced by gravity and it can be
written as $\alpha_{G}=\left(  \pi^{2}a\mu/2\varepsilon e\right)  \alpha_{E}$,
where $\alpha_{E}$ is the usual electric polarizability of the atom induced by
electric fields. In order to make a comparison between the polarization
induced by gravity and that induced by an electric field, let us consider the
equivalent electric field $\mathbf{E}_{eqv}\equiv\mu/e\mathbf{A}$, which is
capable to produce the same acceleration $\mathbf{A}$ in a particle with
charge $e$ and mass $\mu.$ In a TeV-brane, the gravity-induced dipole will be
approximately $10^{9}$ $(\simeq a/\varepsilon)$ greater than the dipole
induced by the equivalent electric field. Despite this impressive
magnification, as the tidal acceleration $\mathbf{A}$ produced by astronomical
bodies will be very tiny, we should expect that gravity-induced dipole will be more significant
in the presence of microscopic black holes.

\subsection{Final remarks}

We would like to mention that in Ref. \cite{fisch}, it was shown that an
effective Hamiltonian of a hydrogen atom, deduced from the covariant Dirac
equation in a curved spacetime up to order $v^{2}/c^{2}$, contains terms that
can mix oppositive-parity state. However, these terms arise as a result of a
relativistic effect in post-Newtonian approximation, while the electric
polarization induced by gravity discussed here is a consequence of the split of
fermions in the extra dimension.

Finally, we should also emphasize that in the fat brane, as electrons and
protons are stuck in different slices of the brane, then every atom should
have an electric dipole with a non-null component in the $z$ direction.
However, we have shown here that the gravitational field of a mass $M$ will
induce a tangential component in the atomic dipole and hence the atom will
produce an electric field in the brane whose pattern can be recognized by 4D-observers.

\end{document}